\documentclass[preprint2]{aastex}
\usepackage{amsmath}
\usepackage{natbib}
\usepackage{rotating}

\begin{document}
\title
{
  Gamma-ray Burst Luminosity Relations:
  Two-dimensional versus Three-dimensional Correlations
}
\author
{
  Bo Yu\altaffilmark{1, 3},
  Shi Qi\altaffilmark{2, 3},
  and
  Tan Lu\altaffilmark{2, 3}
}
\altaffiltext{1}
{
  Department of Physics, Nanjing University, Nanjing 210093, China
}
\altaffiltext{2}
{
  Purple Mountain Observatory, Chinese Academy of Sciences, Nanjing
  210008, China
}
\altaffiltext{3}
{
  Joint Center for Particle, Nuclear Physics and Cosmology, Nanjing
  University --- Purple Mountain Observatory, Nanjing  210093, China
}

\begin{abstract}
  The large scatters of luminosity relations of gamma-ray bursts
  (GRBs) have been one of the most important reasons that prevent the
  extensive applications of GRBs in cosmology.
  In this paper, we extend the two-dimensional (2D) luminosity
  relations with
  $\tau_{\mathrm{lag}}$, $V$, $E_{\mathrm{peak}}$, and
  $\tau_{\mathrm{RT}}$ as the luminosity indicators to three
  dimensions (3D) using the
  same set of luminosity indicators to explore the possibility of
  decreasing the intrinsic scatters.
  We find that, for the 3D luminosity relations between the luminosity
  and an energy scale ($E_{\mathrm{peak}}$) and a time scale
  ($\tau_{\mathrm{lag}}$ or $\tau_{\mathrm{RT}}$), their intrinsic
  scatters are considerably smaller than those of corresponding 2D
  luminosity relations.
  Enlightened by the result and the definition of the luminosity
  (energy released in units of time), we discussed possible reasons
  behind, which may give us helpful suggestions on seeking more
  precise luminosity relations for GRBs in the future.
\end{abstract}


\section{Introduction}

Gamma-ray bursts (GRBs) have recently attracted much attention in
their cosmological applications as the most luminous astrophysical
events observed today.
Based on the correlations between the luminosity/energy and the
measurable parameters of light curves and/or spectra, GRBs can be
used as standard candles after
calibration~\citep[see, for example,][etc.]{Dai:2004tq,
  Ghirlanda:2004fs, Firmani:2005gs, Ghirlanda:2006ax, Schaefer:2006pa,
  Amati:2008hq, Basilakos:2008tp}.
The advantage of the GRBs is their high redshifts due to their high
luminosities.
The $69$ GRBs compiled in~\citet{Schaefer:2006pa} extend the redshift
to $z>6$.
Recently observed GRB 090423 has a redshift of
$z=8.3$~\citep{Tanvir:2009ex, Salvaterra:2009ey}.
However, GRBs are not as ideal standard candles as type Ia supernovae
(SNe Ia).
Compared with SNe Ia, GRBs suffer from the circularity problem due to
the lack of low-redshift samples and the scatters of known luminosity
relations of GRBs are still very large.
There are a few ways proposed in literatures to avoid the circularity
problem.
One can simply fit the calibration parameters and cosmological
parameters simultaneously~\citep{Li:2007re, Qi:2008zk};
and in~\citet{Wang:2008vja}, GRB data are summarized by a set of
model-independent distance measurements;
calibrating GRBs using SNe Ia in their overlapping redshift range was
also proposed~\citep{Kodama:2008dq, Liang:2008kx}
and adopted~\citep{Wei:2008kq, Cardone:2009mr} in cosmological studies
using GRBs.

Till now in most works of cosmological studies using GRBs,
two-dimensional (2D)
luminosity relations are used, which have least calibration
parameters.
For example $\tau_{\mathrm{lag}}$--$L$~\citep{Norris:1999ni},
$V$--$L$~\citep[][
there exist several definitions of $V$, mainly depending on the
smoothing time intervals the reference curve is built upon, and on the
normalization as well]
{Fenimore:2000vs, Reichart:2000kq},
$E_{\mathrm{peak}}$--$E_{\gamma, \mathrm{iso}}$~\citep{Amati:2002ny},
$E_{\mathrm{peak}}$--$E_{\gamma}$~\citep{Ghirlanda:2004me},
$E_{\mathrm{peak}}$--$L$~\citep{Schaefer:2002tf},
and
$\tau_{\mathrm{RT}}$--$L$~\citep{Schaefer:2006pa}
relations.
However, the intrinsic scatters of 2D luminosity relations
are usually very large, which may imply hidden parameters considering
the complication of GRBs.
There are already works which explored the possibility of a
three-dimensional (3D) correlation with negligible scatter.
For example, \citet{Firmani:2006gw} claimed that a temporal parameter
of the prompt emission, the $T_{0.45}$, could reduce the scatter of
the correlation of $L_{\mathrm{iso}}$--$E_{\mathrm{peak}}$ to a
negligible value.
But it was later found that the new proposed relation does not appear
to be as tight as it seemed to
be~\citep{Rossi:2008mv, Collazzi:2008gu}.
In this paper, we extend the luminosity relations used
in~\citet{Schaefer:2006pa} from 2D to 3D and explore the possibility
of decreasing the scatters in the correlations.

\section{Methodology}

In~\citet{Schaefer:2006pa}, five luminosity relations of GRBs as
follows are used:
\begin{eqnarray}
  \label{eq:GRB-lag-L}
  \log \frac{L}{1 \; \mathrm{erg} \; \mathrm{s}^{-1}}
  &=& a_1+b_1 \log
  \left[
    \frac{\tau_{\mathrm{lag}}(1+z)^{-1}}{0.1\;\mathrm{s}}
  \right]
  ,
  \\
  \label{eq:GRB-V-L}
  \log \frac{L}{1 \; \mathrm{erg} \; \mathrm{s}^{-1}}
  &=& a_2+b_2 \log
  \left[
    \frac{V(1+z)}{0.02}
  \right]
  ,
  \\
  \label{eq:GRB-E_peak-L}
  \log \frac{L}{1 \; \mathrm{erg} \; \mathrm{s}^{-1}}
  &=& a_3+b_3 \log
  \left[
    \frac{E_{\mathrm{peak}}(1+z)}{300\;\mathrm{keV}}
  \right]
  ,
  \\
  \label{eq:GRB-E_peak-E_gamma}
  \log \frac{E_{\gamma}}{1\;\mathrm{erg}}
  &=& a_4+b_4 \log
  \left[
    \frac{E_{\mathrm{peak}}(1+z)}{300\;\mathrm{keV}}
  \right]
  ,
  \\
  \label{eq:GRB-tau_RT-L}
  \log \frac{L}{1 \; \mathrm{erg} \; \mathrm{s}^{-1}}
  &=& a_5+b_5 \log
  \left[
    \frac{\tau_{\mathrm{RT}}(1+z)^{-1}}{0.1\;\mathrm{s}}
  \right]
  ,
\end{eqnarray}
where the luminosity $L$ and the total collimation-corrected energy
$E_{\gamma}$ of GRBs are derived respectively from the bolometric peak
flux $P_{\mathrm{bolo}}$ and the bolometric fluence $S_{\mathrm{bolo}}$ of
GRBs through
\begin{eqnarray}
  \label{eq:GRB-L-P_bolo}
  L&=&4\pi d_L^2 P_{\mathrm{bolo}}
  ,
  \\
  \label{eq:GRB-E_gamma-S_bolo}
  E_{\gamma}&=& E_{\gamma,\mathrm{iso}}F_{\mathrm{beam}}
  =4\pi d_L^2 S_{\mathrm{bolo}} (1+z)^{-1} F_{\mathrm{beam}}
  .
\end{eqnarray}
Here $d_L$ is the luminosity distance, which depends on the
cosmological model and is inversely proportional to the value of
Hubble parameter of today.
In this paper, we have adopted the flat $\Lambda$CDM model with
$\Omega_m=0.27$ and in the calculation, we actually replace $d_L$ with
$\bar{d}_L = \frac{H_0}{c} d_L \times 1 \, \mathrm{cm}$ in
Eq.~(\ref{eq:GRB-L-P_bolo}) and
Eq.~(\ref{eq:GRB-E_gamma-S_bolo}), so that the dependence on Hubble
constant is absorbed into the intercepts of the linear luminosity
relations.
For later convenience, we denote these luminosity relations by
\begin{equation}
  \label{eq:GRB-2D}
  y^{(i)} = c_0^{(i, \, i)} + c_1^{(i, \, i)} x^{(i)}
  ,
\end{equation}
where
\begin{eqnarray}
  x^{(1)}
  &=&
  \log
  \left[
    \frac{\tau_{\mathrm{lag}}(1+z)^{-1}}{0.1\;\mathrm{s}}
  \right]
  ,
  \\
  x^{(2)}
  &=&
  \log
  \left[
    \frac{V(1+z)}{0.02}
  \right]
  ,
  \\
  \label{eq:x34}
  x^{(3)}
  &=&
  x^{(4)} =
  \log
  \left[
    \frac{E_{\mathrm{peak}}(1+z)}{300\;\mathrm{keV}}
  \right]
  ,
  \\
  x^{(5)}
  &=&
  \log
  \left[
    \frac{\tau_{\mathrm{RT}}(1+z)^{-1}}{0.1\;\mathrm{s}}
  \right]
  ,
  \\
  \label{eq:y1235}
  y^{(1)}
  &=&
  y^{(2)} = y^{(3)} = y^{(5)} =
  \log \frac{L}{1 \; \mathrm{erg} \; \mathrm{s}^{-1}}
  ,
  \\
  \label{eq:y4}
  y^{(4)}
  &=&
  \log \frac{E_{\gamma}}{1\;\mathrm{erg}}
  ,
\end{eqnarray}
and
\begin{equation}
  c_0^{(i, \, i)} = a_i, \quad c_1^{(i, \, i)} = b_i
  .
\end{equation}
The coefficient $c$ is given two superscripts to incorporate 3D
correlations introduced just below.
Hereafter, we denote a luminosity relation by the superscript pair of
the corresponding $c$.

The above 2D luminosity relations connect measurable parameters of
light curves and/or spectra with GRB luminosity/energy.
They are empirical though there exist some explanations
(see~\citet{Schaefer:2006pa} and the references therein).
Since the physical processes of GRBs are very complicated, the real
correlations between GRB luminosity/energy and the parameters of light
curves and/or spectra, generally speaking, should be more complicated
than the above simple relations.
The large scatters in the above luminosity relations also imply that
there may be hidden parameters not included.
Motivated by this, we extend the 2D luminosity relations to 3D and
investigate if there are any improvements.
See Eq.~(\ref{eq:x34}), Eq.~(\ref{eq:y1235}), and Eq.~(\ref{eq:y4}),
since $x^{(3)} = x^{(4)}$ and $y^{(4)}$ is different from
$y^{(i)}$ ($i = 1, \, 2, \, 3, \, 5$),
we introduce 3D correlations as follows.
For $(i, \, j)$ with both $i$ and $j$ in $(1, \, 2, \, 3, \, 5)$ and
$i < j$, we introduce 3D correlations as
\begin{equation}
  \label{eq:GRB-3D_1}
  y^{(i)} = c_0^{(i, \, j)}
  + c_1^{(i, \, j)} x^{(i)}
  + c_2^{(i, \, j)} x^{(j)}
  ,
\end{equation}
for the case of $(i, \, j) = (3, \, 4)$, the 3D
correlation is defined as
\begin{equation}
  \label{eq:GRB-3D_2}
  y^{(3)} = c_0^{(3, \, 4)}
  + c_1^{(3, \, 4)} x^{(3)}
  + c_2^{(3, \, 4)} y^{(4)}
  ,
\end{equation}
and for
$(i, \, j) = (1, \, 4)$, $(2, \, 4)$, or $(4, \, 5)$
\begin{equation}
  \label{eq:GRB-3D_3}
  y^{(4)} = c_0^{(i, \, j)}
  + c_1^{(i, \, j)} x^{(i)}
  + c_2^{(i, \, j)} x^{(j)}
  .
\end{equation}
The luminosity relation $(3, \, 4)$ is different from other 3D
luminosity relations in that there is only one luminosity indicator
(i.e. $E_{\mathrm{peak}}$) in it, while there are two in other 3D
luminosity relations.
The luminosity relations in Eq.~(\ref{eq:GRB-3D_1}) and those in
Eq.~(\ref{eq:GRB-3D_3}) are different from each other in that the
left-hand side of Eq.~(\ref{eq:GRB-3D_1}) is
$\log \frac{L}{1 \; \mathrm{erg} \; \mathrm{s}^{-1}}$,
while the left-hand side of Eq.~(\ref{eq:GRB-3D_3}) is
$\log \frac{E_{\gamma}}{1\;\mathrm{erg}}$.

We investigate the quality of the luminosity relations mainly by
comparing their intrinsic scatters.
Since we are free to multiply the luminosity relations by a constant
when introducing 3D correlations and such a factor would increase the
intrinsic scatter by the same multiple as the factor, we need to
normalize the relations in order to compare the intrinsic scatters.
Bearing in mind that one of the most important purposes for exploring
the luminosity relations is to use them in distance measurements,
what we do here is just dividing Eq.~(\ref{eq:GRB-3D_2})
by a factor of $1 - c_2^{(3, \, 4)}$
so that $\log (d_L)$ have the same coefficient in all luminosity
relations above.

Comparing the 3D luminosity relations with the 2Ds, as an example,
comparing Eq.~(\ref{eq:GRB-3D_1}) with Eq.~(\ref{eq:GRB-2D}), one can
see that, by introducing the 3D correlations in this way, we actually
treat $x^{(j)}$ as the hidden parameter of the 2D correlation of
Eq.~(\ref{eq:GRB-2D}) and write it explicitly in the 3D correlation
of Eq.~(\ref{eq:GRB-3D_1}).
So, in addition to examine the quality of the luminosity relations by
comparing their intrinsic scatters, we also calculated the correlation
coefficients between the residual of the fit of 2D correlations
and the possible hidden parameters we introduced to extend
correlations from 2D to 3D.

In the fit of the luminosity relations,
we used the techniques presented in~\citet{Agostini:2005fe}, following
which the likelihood function for the coefficients $c$ and the
intrinsic scatter is (for the cases of Eq.~(\ref{eq:GRB-3D_1}). The
other cases of 3D luminosity relations are similar and the likelihood
function for the 2D luminosity relations can be obtained just by
setting $c_2 = 0$)
\begin{align}
  \label{eq:likelihood}
  \mathcal{L}(c, \, \sigma_{\mathrm{int}})
  &\propto \prod_k
  \frac{1}
  {
    \sqrt
    {
      \sigma_{\mathrm{int}}^2
      + \sigma_{y_k^{(i)}}^2
      + c_1^2 \sigma_{x_k^{(i)}}^2
      + c_2^2 \sigma_{x_k^{(j)}}^2
    }
  }
  \nonumber \\
  & 
  \times
  \exp
  \left[
    -
    \frac
    {
      \left(
        y_k^{(i)}
        - c_0
        - c_1 x_k^{(i)}
        - c_2 x_k^{(j)}
      \right)^2
    }
    {
      2
      \left(
        \sigma_{\mathrm{int}}^2
        + \sigma_{y_k^{(i)}}^2
        + c_1^2 \sigma_{x_k^{(i)}}^2
        + c_2^2 \sigma_{x_k^{(j)}}^2
      \right)
    }
  \right]
  ,
\end{align}
where $k$ runs over GRBs with corresponding quantities available.
In the calculation, Markov chain Monte Carlo techniques are used.
For each luminosity relations, A Markov chain with samples of order
$10^6$ is generated according to the likelihood function and then
properly burned in and thinned to derive statistics of interested
parameters.

For the GRB data, we used the compilation in~\citet{Schaefer:2006pa},
which includes 69 GRBs.
When considering error propagation from a quantity, say $\xi$ with
error $\sigma_{\xi}$, to its logarithm, we set
$
\frac
{
  \log(\xi + \sigma_{\xi}^{+})
  +
  \log(\xi - \sigma_{\xi}^{-})
}
{
  2
}
$
and
$
\frac
{
  \log(\xi + \sigma_{\xi}^{+})
  -
  \log(\xi - \sigma_{\xi}^{-})
}
{
  2
}
$
as the center value and the error of the logarithm correspondingly.
This requires $\xi > \sigma_{\xi}^{-}$ (the quantities we are
interested in here are all positive).
Due to the limitation of the data,
for a given luminosity relation $(i, \, j)$, not all the GRBs have all
of the needed observational quantities available and satisfy
$\xi > \sigma_{\xi}^{-}$ at the same time. By set $(i, \, j)$
we denote the maximum GRB set that can be used in the luminosity
relation $(i, \, j)$.
The numbers of GRBs of different sets are presented in
Table~\ref{tab:fit}.

\section{Results and discussion}

\begin{sidewaystable}[htbp]
  \centering
  \tiny
  \begin{tabular}{|c|c|c|c|c|c|}
    \hline
    (i, j) & 1 & 2 & 3 & 4 & 5 \\
    \hline
    1
    &
    \begin{tabular}{@{} c @{}}
      $32$ \\
      $(-3.994_{-0.077}^{+0.078}, \, -0.79_{-0.11}^{+0.11})$ \\
      $0.404_{-0.055}^{+0.067}$
    \end{tabular}
    &
    \begin{tabular}{@{} c @{}}
      $22$ \\
      $(-3.96_{-0.10}^{+0.10}, \, -0.68_{-0.18}^{+0.19}, \, 0.51_{-0.36}^{+0.38})$ \\
      $0.414_{-0.067}^{+0.087}$ \\
      $[-0.01_{-0.10}^{+0.09}, \, 0.09_{-0.10}^{+0.09}]$
    \end{tabular}
    &
    \begin{tabular}{@{} c @{}}
      $30$ \\
      $(-4.013_{-0.060}^{+0.059}, \, -0.618_{-0.091}^{+0.091}, \, 0.80_{-0.16}^{+0.16})$ \\
      $0.279_{-0.042}^{+0.052}$ \\
      $[0.124_{-0.076}^{+0.079}, \, 0.142_{-0.066}^{+0.064}]$
    \end{tabular}
    &
    \begin{tabular}{@{} c @{}}
      $13$ \\
      $(-5.575_{-0.083}^{+0.074}, \, 0.08_{-0.14}^{+0.14}, \, 1.44_{-0.18}^{+0.17})$ \\
      $0.15_{-0.08}^{+0.11}$ \\
      $[0.01_{-0.11}^{+0.10}]$
    \end{tabular}
    &
    \begin{tabular}{@{} c @{}}
      $31$ \\
      $(-3.952_{-0.081}^{+0.081}, \, -0.60_{-0.14}^{+0.14}, \, -0.31_{-0.17}^{+0.17})$ \\
      $0.344_{-0.051}^{+0.063}$ \\
      $[0.059_{-0.083}^{+0.084}, \, 0.111_{-0.076}^{+0.072}]$
    \end{tabular}
    \\
    \hline
    2 & - &
    \begin{tabular}{@{} c @{}}
      $44$ \\
      $(-3.712_{-0.081}^{+0.080}, \, 1.02_{-0.22}^{+0.23})$ \\
      $0.508_{-0.055}^{+0.066}$
    \end{tabular}
    &
    \begin{tabular}{@{} c @{}}
      $42$ \\
      $(-3.875_{-0.067}^{+0.067}, \, 0.65_{-0.19}^{+0.19}, \, 1.09_{-0.19}^{+0.19})$ \\
      $0.366_{-0.043}^{+0.051}$ \\
      $[0.141_{-0.075}^{+0.078}, \, 0.055_{-0.065}^{+0.064}]$
    \end{tabular}
    &
    \begin{tabular}{@{} c @{}}
      $22$ \\
      $(-5.652_{-0.057}^{+0.052}, \, 0.15_{-0.22}^{+0.23}, \, 1.56_{-0.15}^{+0.15})$ \\
      $0.178_{-0.051}^{+0.064}$ \\
      $[-0.020_{-0.078}^{+0.075}]$
    \end{tabular}
    &
    \begin{tabular}{@{} c @{}}
      $41$ \\
      $(-3.702_{-0.071}^{+0.071}, \, 0.46_{-0.23}^{+0.22}, \, -0.65_{-0.16}^{+0.16})$ \\
      $0.432_{-0.050}^{+0.060}$ \\
      $[0.075_{-0.081}^{+0.082}, \, 0.023_{-0.074}^{+0.071}]$
    \end{tabular}
    \\
    \hline
    3 & - & - &
    \begin{tabular}{@{} c @{}}
      $63$ \\
      $(-3.999_{-0.058}^{+0.058}, \, 1.37_{-0.12}^{+0.12})$ \\
      $0.422_{-0.041}^{+0.048}$
    \end{tabular}
    &
    \begin{tabular}{@{} c @{}}
      $27$ \\
      $(-3.3_{-1.6}^{+1.6}, \, 0.91_{-0.43}^{+0.48}, \, 0.12_{-0.29}^{+0.27})$ \\
      $0.50_{-0.14}^{+0.26}$ \\
      $[-0.08_{-0.26}^{+0.14}, \, -0.34_{-0.26}^{+0.15}]$
    \end{tabular}
    &
    \begin{tabular}{@{} c @{}}
      $56$ \\
      $(-3.852_{-0.048}^{+0.048}, \, 0.90_{-0.11}^{+0.11}, \, -0.624_{-0.090}^{+0.089})$ \\
      $0.293_{-0.032}^{+0.037}$ \\
      $[0.128_{-0.055}^{+0.057}, \, 0.162_{-0.057}^{+0.060}]$
    \end{tabular}
    \\
    \hline
    4 & - & - & - &
    \begin{tabular}{@{} c @{}}
      $27$ \\
      $(-5.626_{-0.047}^{+0.044}, \, 1.51_{-0.11}^{+0.11})$ \\
      $0.159_{-0.046}^{+0.054}$
    \end{tabular}
    &
    \begin{tabular}{@{} c @{}}
      $23$ \\
      $(-5.661_{-0.059}^{+0.055}, \, 1.56_{-0.14}^{+0.15}, \, 0.00_{-0.12}^{+0.11})$ \\
      $0.176_{-0.052}^{+0.064}$ \\
      $[-0.017_{-0.078}^{+0.075}]$
    \end{tabular}
    \\
    \hline
    5 & - & - & - & - &
    \begin{tabular}{@{} c @{}}
      $61$ \\
      $(-3.766_{-0.065}^{+0.065}, \, -0.88_{-0.11}^{+0.11})$ \\
      $0.456_{-0.044}^{+0.051}$
    \end{tabular}
    \\
    \hline
  \end{tabular}
  \caption
  {
    Fit of 2D and 3D luminosity relations.
    In every grid, the first row is the number of GRBs of
    set $(i, \, j)$, 
    the vector below enclosed by parentheses is the vector of
    $c$ for the luminosity relation $(i, \, j)$, and what follows
    next is the intrinsic scatter.
    For 3D luminosity relations, their reduction in the intrinsic
    scatters compared to corresponding 2D luminosity relations are
    presented in the brackets:
    for 3D luminosity relations in Eq.~(\ref{eq:GRB-3D_1}) and
    Eq.~(\ref{eq:GRB-3D_2}), the reduction corresponds to the 2D
    luminosity relation $(i, \, i)$ and $(j, \, j)$ in turn and for
    those in Eq.~(\ref{eq:GRB-3D_3}), corresponds to $(4, \, 4)$.
    The statistics in the table are for the median values and the
    errors of $1 \sigma$ ($68.3\%$) confidence level.
  }
  \label{tab:fit}
\end{sidewaystable}

\begin{sidewaystable}[htbp]
  \centering
  \begin{tabular}{|c|c|c|c|c|c|c|}
    \hline
    & $x^{(1)}$ & $x^{(2)}$ & $x^{(3)}$ & $x^{(5)}$
    & $y^{(3)}$ & $y^{(4)}$ \\
    \hline
    $(1, \, 1)$
    & - & $0.31_{-0.14}^{+0.12}$ & $0.712_{-0.077}^{+0.053}$ & $-0.18_{-0.14}^{+0.15}$
    & - & - \\
    \hline
    $(2, \, 2)$
    & $-0.647_{-0.069}^{+0.093}$ & - & $0.689_{-0.060}^{+0.042}$ & $-0.474_{-0.081}^{+0.097}$
    & - & - \\
    \hline
    $(3, \, 3)$
    & $-0.732_{-0.040}^{+0.046}$ & $0.472_{-0.044}^{+0.038}$ & - & $-0.634_{-0.044}^{+0.050}$
    & - & $-0.21_{-0.09}^{+0.10}$ \\
    \hline
    $(4, \, 4)$
    & $0.223_{-0.012}^{+0.008}$ & $0.380_{-0.020}^{+0.014}$ & - & $-0.007_{-0.037}^{+0.037}$
    & $-0.09_{-0.11}^{+0.11}$ & - \\
    \hline
    $(5, \, 5)$
    & $-0.36_{-0.10}^{+0.10}$ & $0.185_{-0.089}^{+0.087}$ & $0.740_{-0.045}^{+0.034}$ & -
    & - & - \\
    \hline
  \end{tabular}
  \caption
  {
    Correlation coefficients between the residuals of the fit of 2D
    correlations and possible hidden parameters.
    The statistics in the table are for the median values and the
    errors of $1 \sigma$ ($68.3\%$) confidence level.
  }
  \label{tab:correlation}
\end{sidewaystable}

\begin{figure}[htbp]
  \centering
  \includegraphics[width = 0.45 \textwidth]{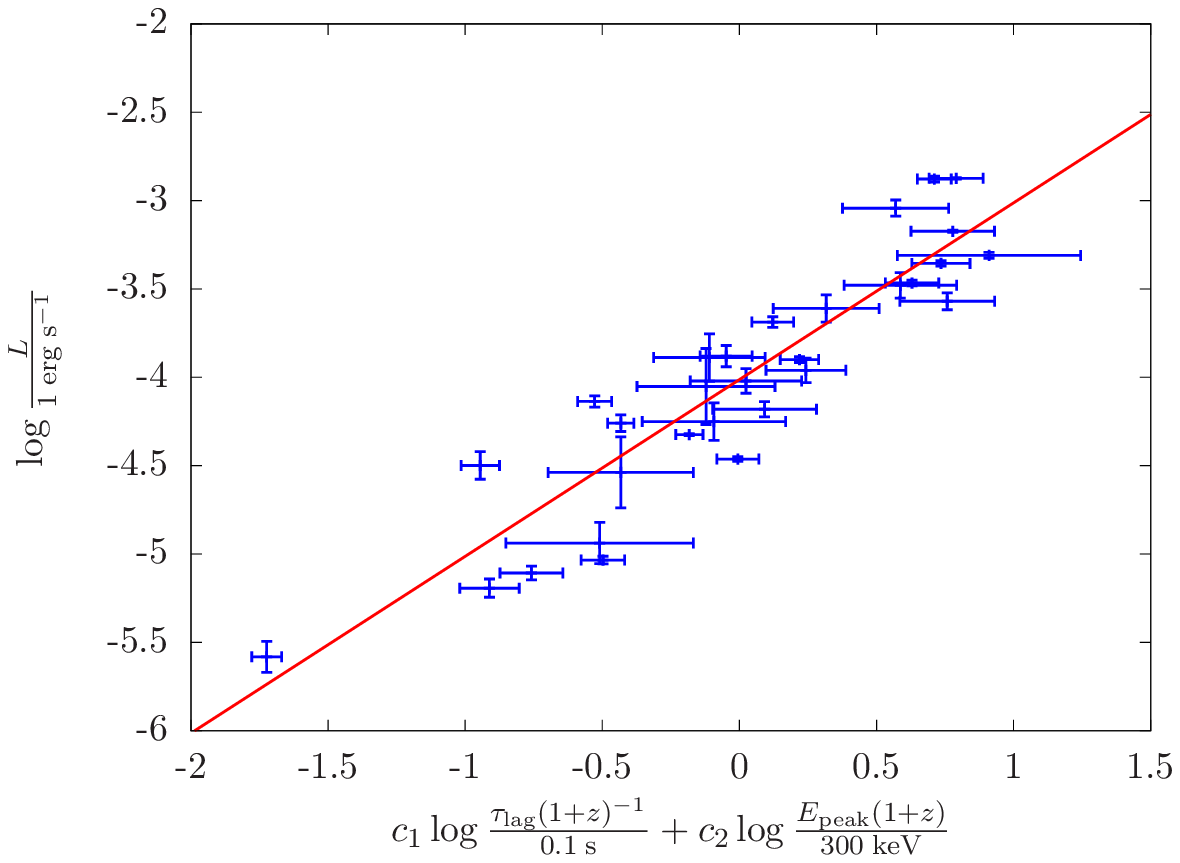}
  \includegraphics[width = 0.45 \textwidth]{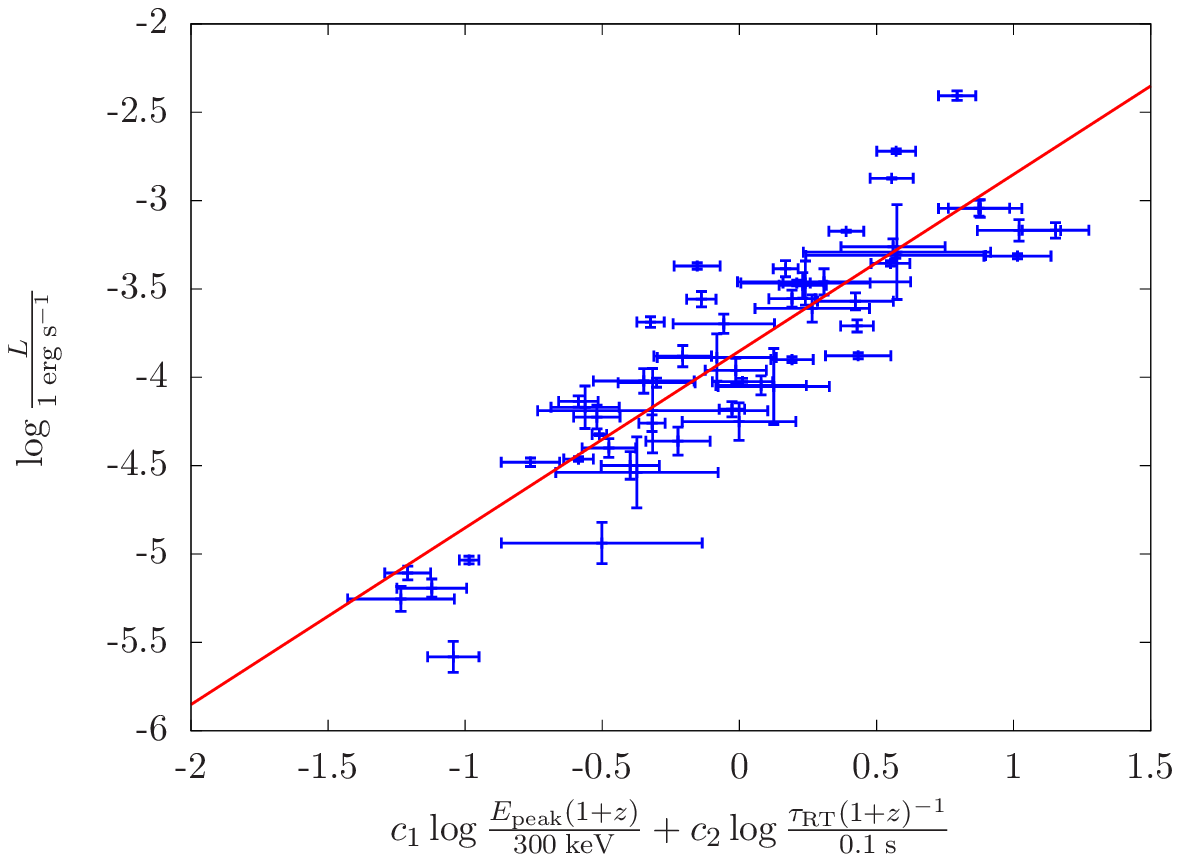}
  \caption
  {
    Plots of the 3D luminosity relation $(1, \, 3)$ and $(3, \, 5)$.
  }
  \label{fig:xy_plot}
\end{figure}

We summarize our results in Tables~\ref{tab:fit}
and~\ref{tab:correlation}.
In Table~\ref{tab:fit}, by comparing the intrinsic scatters of the 3D
luminosity relations with those of the corresponding 2D luminosity
relations, we find that only for the cases of
$(1, \, 3)$ and $(3, \, 5)$
the intrinsic scatters of the 3D luminosity relations are considerably
smaller than those of their corresponding 2D luminosity relations,
and for all other cases, the intrinsic scatters of the 3D luminosity
relations are either very close to the smaller one of the intrinsic
scatters of the corresponding 2D luminosity relations (for
correlations in Eq.~(\ref{eq:GRB-3D_3}), close to the intrinsic
scatter of luminosity relation $(4, \, 4)$) or even greater than those
of the corresponding 2D luminosity relations
(for the case $(3, \, 4)$).
The correlations presented in Table~\ref{tab:correlation} also give
the same implication.
Only the correlation coefficients between the residual of fit of
$(1, \, 1)$ and $x^{(3)}$,
$(3, \, 3)$ and $x^{(1)}$,
$(5, \, 5)$ and $x^{(3)}$
are approximately greater than $0.7$.
We presented the plots of the luminosity relation
$(1, \, 3)$ and $(3, \, 5)$
in Figure~\ref{fig:xy_plot}.

It is interesting to note that the 3D luminosity relations that have
considerable improvement in the intrinsic scatter are the luminosity
relations in Eq.~(\ref{eq:GRB-3D_1}), with one of the luminosity
indicators being an energy scale ($E_{\mathrm{peak}}$) and the other a
time scale ($\tau_{\mathrm{lag}}$ or $\tau_{\mathrm{RT}}$).
Enlightened by this, one may naturally guess that probably the
luminosities of GRBs are mainly determined by a characteristic energy
scale and a characteristic time scale and $E_{\mathrm{peak}}$ is
correlated to the characteristic energy scale, $\tau_{\mathrm{lag}}$
and $\tau_{\mathrm{RT}}$ to the characteristic time scale, so that 3D
luminosity relations with $E_{\mathrm{peak}}$ and $\tau_{\mathrm{lag}}$
or $\tau_{\mathrm{RT}}$ as luminosity indicators could significantly
reduce the intrinsic scatters compared to corresponding 2D luminosity
relations.
This is easy to understand, since the definition of the luminosity is
the amount of the energy released in units of time. If one knows the
released energy and the time duration of a GRB, the (averaged)
luminosity can be calculated immediately.
Obviously, the 2D correlation between the luminosity and the
energy or the time duration would not be complete unless the time
duration or the energy is constant for all samples or there is a
strong correlation between the energy and the time duration.
This, if proven true, also explains why the 2D luminosity relations
$(3, \, 3)$ and $(4, \, 4)$
have little correlation with each other despite their similarity at
first sight
(see discussion in Section 4.7 of~\citet{Schaefer:2006pa};
the correlation coefficient between the residual of fit of
$(3, \, 3)$ and $y^{(4)}$
and the correlation coefficient between the residual of fit of
$(4, \, 4)$ and $y^{(3)}$ also show their weak correlation).
This is because of the difference between $E_{\gamma}$ and $L$, where
a new independent variable---the characteristic time scale---enters.
In fact, we could check the guess partially with current data by
examining the correlation between $x^{(i)}$s.
In Table~\ref{tab:cor_x}, one can find that $x^{(1)}$, $x^{(2)}$, and
$x^{(5)}$ are strongly correlated with each other, while all of them
have only weak correlation with $x^{(3)}$
(see also~\citet{Schaefer:2001ap} and Section 4.5
of~\citet{Schaefer:2006pa} for discussions on the correlations between
the luminosity indicators).
This is consistent with the guess that the luminosity indicators
$\tau_{\mathrm{lag}}$ and $\tau_{\mathrm{RT}}$ are correlated
with a characteristic time scale and $E_{\mathrm{peak}}$ is correlated
with a characteristic energy scale which is independent of the
characteristic time scale.
In addition, the correlations presented in Table~\ref{tab:cor_x} seems
to imply that $V$ is also correlated with the characteristic time
scale.
Accordingly, there is indeed some reduction in the intrinsic
scatter for the 3D luminosity relation $(2, \, 3)$ compared with
corresponding 2D luminosity relations, but such reduction is
relatively smaller than that of $(1, \, 3)$ and $(3, \, 5)$.
Maybe the correlation of $V$ with the characteristic time scale is not
so strong as that of $\tau_{\mathrm{lag}}$ or $\tau_{\mathrm{RT}}$.
\begin{table}[htbp]
  \centering
  \begin{tabular}{|c|c|c|c|c|}
    \hline
    &          $x^{(1)}$ & $x^{(2)}$ & $x^{(3)}$ & $x^{(5)}$ \\
    \hline
    $x^{(1)}$ & 1.0      & -0.73    & -0.35    & 0.72 \\
    \hline
    $x^{(2)}$ & -0.73    & 1.0      & 0.41     & -0.63 \\
    \hline
    $x^{(3)}$ & -0.35    & 0.41     & 1.0      & -0.30 \\
    \hline
    $x^{(5)}$ & 0.72     & -0.63    & -0.30    & 1.0 \\
    \hline
  \end{tabular}
  \caption
  {
    Correlation coefficients between $x^{(1)}$, $x^{(2)}$, $x^{(3)}$,
    and $x^{(5)}$.
  }
  \label{tab:cor_x}
\end{table}

If our guess about the GRB luminosity relations is correct, it would
be very enlightening. It suggests us to include an energy scale and
a corresponding time scale for seeking more precise luminosity
relations for GRBs in the future.
However, it should be emphasized that, even if it is true, only
appropriate energy and time scales might significantly reduce the
intrinsic scatter bearing in mind the situation of some 3D
luminosity relations~\citep[see, for example,][]{Firmani:2006gw,
  Rossi:2008mv, Collazzi:2008gu}.

\section{Summary}

In this paper, we extend the widely used 2D luminosity relations to 3D
by using the same set of luminosity indicators,
i.e. $\tau_{\mathrm{lag}}$, $V$, $E_{\mathrm{peak}}$, and
$\tau_{\mathrm{RT}}$,
and check the improvement in the quality of the luminosity relations.
We find that, for the 3D luminosity relations between the luminosity
and an energy scale ($E_{\mathrm{peak}}$) and a time scale
($\tau_{\mathrm{lag}}$ or $\tau_{\mathrm{RT}}$), their intrinsic
scatters are considerably smaller than those of corresponding 2D
luminosity relations.
The correlations between the residuals of fit of the 2D luminosity
relations and the luminosity indicators also give the same
implication.
Enlightened by the result and the definition of the luminosity
(energy released in units of time), we discussed possible reasons
behind, which may give us helpful suggestions on seeking more
precise luminosity relations for GRBs in the future.

\acknowledgments

We thank the anonymous referee for detailed and helpful comments and
suggestions.
This research was supported by the National Natural Science Foundation
of China under Grant No.~10973039 and Jiangsu Planned Projects for
Postdoctoral Research Funds 0901059C (for Shi Qi).

\end{document}